\documentstyle[psfig,a4]{article}

\begin{document}

\renewcommand{\thesection}{\arabic{section}} \renewcommand{\figurename}{%
\small Fig.} \renewcommand{\theequation}{\arabic{section}.\arabic{equation}}

\begin{flushleft}

{\small 
 Physica A (to appear)}
\vspace{2cm}

{
\Large\bf
Boundary and finite-size effects in small magnetic systems
}      

\vspace{0.1cm}

\end{flushleft}

\begin{flushright}
\parbox[t]{11cm}{

H Kachkachi$^*$ and D A Garanin$^\dagger$
\vspace{0.1cm}

\small

$^*$Laboratoire de Magn\'{e}tisme et d'Optique, Universit\'e de Versailles St.
Quentin, 
45 av. des Etats-Unis, 78035 Versailles, France\\
\vspace{-0.4cm}

$^\dagger$Max-Planck-Institut f\"ur Physik komplexer Systeme, N\"othnitzer Strasse 38,
D-01187 Dresden, Germany\\ 
\vspace{0.6cm}

{\bf Abstract.} \hspace{1mm}
We study the effect of free boundaries in finite magnetic systems 
of cubic shape on the field and temperature dependence of the magnetisation 
within the isotropic model of $D$-component spin vectors in the limit $D\to
\infty$. 
This model is described by a closed system of equations and captures the
Goldstone-mode effects such as global rotation of the magnetic moment and
spin-wave fluctuations.  
We have obtained an exact relation between the intrinsic (short-range) 
magnetisation $M=M(H,T)$ of the system
and the induced magnetisation $m=m(H,T)$ which is induced by the field.
We have shown, analytically at low temperatures and fields and numerically in a wide
range of these parameters, that boundary effects leading to the decrease of $M$ with respect to
the bulk value are stronger than the finite-size effects rendering a positive 
contribution to $M$.  
The inhomogeneities of the magnetisation caused by the boundaries are long
ranged and extend far into the depth of the system.
} 
\vspace{0.8cm}

\end{flushright}

\renewcommand{\thefootnote}{\fnsymbol{footnote}} \renewcommand{%
\footnoterule}{\rule{0cm}{0cm}} 
\footnotetext[1]{
kachkach@physique.uvsq.fr} 
\footnotetext[2]{
www.mpipks-dresden.mpg.de/$\sim$garanin/; garanin@mpipks-dresden.mpg.de}

\section{Introduction}

\setcounter{equation}{0}

Small magnetic particles have been of much interest owing to their
technological applications, mainly in the area of information storage. From
the experimental and theoretical point of view, these systems are very
interesting for they show superparamagnetism at high temperature and
exponentially slow relaxation rates at low temperature due to anisotropy
barriers. Although in most of theoretical approaches to the dynamics of a
small magnetic particle the latter is considered as a single magnetic
moment, deviations from this simple picture become crucial with the
reduction of the system size. In magnetic {\em nanoparticles} (see, for a
review, Ref.\ \cite{dorfiotro97}), the contribution of the surface to the
thermodynamic properties becomes comparable with the bulk contribution, and
the magnetisation and other characteristics may be spatially inhomogeneous.
The latter is realised due to additional thermal disordering near the
surface at elevated temperatures or due to the breaking of symmetry of the
crystal field for surface spins, which may result in a strong surface
anisotropy. Another manifestation of the symmetry breaking at the surface is
the possible unquenching of the orbital moments of surface spins, which may
be responsible for a significant increase of the particle's magnetic moment,
e.g. in 3$d$ elements \cite{erietal92}.

A finite-size magnetic system with realistic free boundary conditions
presents a spati\-ally inhomogeneous many-body problem. A great deal of work
up to date has been based on the Monte Carlo (MC) technique for the Ising
model. MC technique was also used with the more adequate classical
Heisenberg model to simulate idealised isotropic nanoparticles with simple
cubic (sc) structure and spherical shape in Ref.\ \cite{wil74}. Magnetic
nanoparticles with realistic lattice structure were simulated in Ref.\ \cite
{kacetal99} taking into account surface anisotropy and dipole-dipole
interaction.

Familiar analytical methods such as the mean-field approximation (MFA) and
spin-wave theory (SWT) are (at least in their standard form) inapproriate
for finite magnetic systems because of the Goldstone mode corresponding to
the global rotation of the magnetic moment in zero field. A more
sophysticated way to analytically treat finite magnets is to take the limit $%
D\to \infty $ for the model of $D$-component classical spin vectors, which
was introduced by Stanley \cite{sta68prl}. Stanley has shown \cite{sta68pr}
that for $D\to \infty $ the partition function for this model in the bulk
coincides with that of the exactly solvable spherical model (SM) \cite
{berkac52}. Both models provide a reasonably good approximation (about 5\%
overall accuracy) to the isotropic classical Heisenberg model in three
dimensions. For spatially inhomogeneous systems such as magnetic particles
or films with free boundary conditions, these two models become
nonequivalent, the SM having been shown to yield unphysical results because
of the inappropriate global spin constraint \cite{barfis73}. This drawback
was fixed in an improved version of the SM using {\em local} spin
constraints on each lattice site \cite{kno73,cosmazmih76}. Yet two profound
differences between the $D\to \infty $ model and the SM cannot be
eliminated. First, there are two (longitudinal and transverse) correlation
functions in the former \cite{gar97zpb} and only one CF in the latter.
Second, generalization for the anisotropic case is only possible for the $%
D\to \infty $ model. The resulting anisotropic spherical model (ASM) was
applied to describe phase transitions in domain walls \cite{gar96jpa} and in
magnetic films \cite{gar96jpal99jpa}. The main feature of the $D\to \infty $
model or the ASM is that it takes into account Goldstone modes in the
system, which prevent phase transitions for isotropic systems in dimensions $%
d\leq 2$. In contrast with the linear SWT, this model self-consistently
generates a gap in the correlation functions which avoids the infrared
divergencies. Anisotropy plays a crucial role in these systems since it
breaks the symmetry of the system and makes phase transitions in two
dimensions possible.

So far, the ASM was only applied to spatially inhomogeneous systems in the
plane geometry \cite{gar96jpa,gar96jpal99jpa,gar98pre}. Here we extend it
for {\em finite} box-shaped magnetic systems with free and periodic boundary
conditions (fbc and pbc), restricting ourselves to the idealised case of
isotropic coupling and simple cubic lattice and reserving consideration of
realistic lattice structures, bulk and surface anisotropies, etc., for the
subsequent work. We obtain analytical asymptotes at low temperatures and
fields and a complete numerical solution in the whole range of $T$ and $H$
for the intrinsic magnetisation $M=M(H,T)$ and the induced magnetisation $%
m=m(H,T)$, and we prove the {\em exact} relation $m=MB({\cal N}MH/T)$
between them, ${\cal N}=N_{x}N_{y}N_{z}$ being the number of sites in the
system and $B(x)$ the Langevin function. For the system with fbc, the value
of $M$ is significantly reduced with respect to that for the pbc model and
to the bulk due to boundary effects. We consider temperature dependences of 
{\em local} magnetisation $M_{i}$ and estimate the critical indices for the
magnetisation at the faces, edges, and corners of the system. We show that
because of the Goldstone modes, the inhomogeneities of $M_{i}$ extend far
into the depth of the particle, i.e., the magnetisation profile in our
isotropic model is long ranged. This feature is in accord with the MC
simulations for the Heisenberg model \cite{wil74,kacetal99}, as opposed to
the MFA predictions \cite{wil74}.

The rest of the paper is organized as follows. In Sec.\ \ref{basic} we give
the equations describing the spatially inhomogeneous ASM {\em in a magnetic
field}, generalizing the results of Ref.\ \cite{gar96jpa}. Further,
restricting ourselves to the isotropic model, we define the two
magnetisations mentioned above and derive relations between them. In Sec.\ 
\ref{pbc} we consider the model with periodic boundary conditions and derive
low-temperature and low-field analytical expressions for the intrinsic
magnetisation. In Sec.\ \ref{LowT} we take into account boundary effects for
the fbc model at low temperatures. In Sec.\ \ref{numres} the numerical
results for the temperature, field, and spatial dependencies of the
magnetisation of finite magnetic systems are presented.

\section{Hamiltonian and basic equations}

\label{basic}

\ We start with the Hamiltonian of the uniaxially anisotropic classical $D$%
-component vector model, which can be written in the form 
\begin{equation}
{\cal H}=-{\bf H\cdot }\sum \limits_{i}{\bf s}_{i}-\frac{1}{2}\sum
\limits_{i,j}J_{ij}\left( s_{zi}s_{zj}+\eta \sum \limits_{\alpha
=2}^{D}s_{\alpha i}s_{\alpha j}\right) ,  \label{DHam}
\end{equation}
where ${\bf s}_{i}$ is the normalized $D$-component vector, $\left| {\bf s}%
_{i}\right| =1$, and $\eta \leq 1$ is the dimensionless anisotropy factor; $%
{\bf H}$ is the magnetic field, and $J_{ij}$ the exchange coupling.

In the mean-field approximation the Curie temperature of this model is $%
T_{c}^{{\rm MFA}}=$ $J_{0}/D$, where $J_{0}$ is the zero Fourier component
of $J_{ij}$. It is convenient to use $T_{c}^{{\rm MFA}}$ as the energy scale
and introduce the dimensionless variables 
\begin{equation}  \label{DefPar}
\theta \equiv T/T_{c}^{{\rm MFA}}, \qquad {\bf h\equiv H/}J_{0}, \qquad
\lambda _{ij}\equiv J_{ij}/J_{0}.
\end{equation}
For the nearest-neighbour (nn) interaction $J_{ij}$ with $z$ neighbours, $%
\lambda _{ij}$ is equal to $1/z$ if sites $i$ and $j$ are nearest neighbours
and zero otherwise.

Using the diagram technique for classical spin systems \cite
{garlut84d,gar94jsp,gar96prb} in the limit $D\rightarrow \infty $ and
generalising the results of Ref.\ \cite{gar96jpa} for spatially
inhomogeneous systems to include the magnetic field ${\bf h=}h_{x}{\bf e}%
_{x}+h_{z}{\bf e}_{z}$, one arrives at the closed system of equations for
the average magnetisation components $\alpha =1$ ($z$) and $\alpha =2$ ($x$)
and correlation functions for the remaining spin components labelled by $%
\alpha \geq 3$, 
\begin{equation}
m_{xi}\equiv \left\langle s_{xi}\right\rangle ,\qquad m_{zi}\equiv
\left\langle s_{zi}\right\rangle ,\qquad s_{ij}\equiv D\left\langle
s_{\alpha i}s_{\alpha j}\right\rangle .  \label{Defms}
\end{equation}
Here all correlation functions are equal to each other by symmetry. The
system of equations describing the anisotropic spherical model consists of
equations for the magnetisation components 
\begin{equation}
m_{xi}=G_{i}(h_{x}+\eta \sum\limits_{j}\lambda _{ij}m_{xj}),\qquad
m_{zi}=G_{i}(h_{z}+\sum\limits_{j}\lambda _{ij}m_{zj}),  \label{MagEq}
\end{equation}
and the Dyson equation for the correlation function 
\begin{equation}
s_{il}=\theta G_{i}\delta _{il}+\eta G_{i}\sum\limits_{j}\lambda
_{ij}s_{jl},  \label{CFEq}
\end{equation}
where $\delta _{il}$ is the Kronecker symbol. $G_{i}$ is the so-called {\it %
gap parameter} to be determined from the set of constraint equations on all
sites $i=1,\ldots ,{\cal N}$ of the lattice 
\begin{equation}
s_{ii}+{\bf m}_{i}^{2}=1,  \label{Constraint}
\end{equation}
where ${\bf m}_{i}^{2}=m_{zi}^{2}+m_{xi}^{2}.$

The system of equations above describing the anisotropic spherical model
(ASM) will be used in its full form elsewhere. In the present work, we will
consider isotropic ($\eta =1$) magnetic systems of box shape of volume $%
{\cal N}=N_{x}N_{y}N_{z}$, with simple-cubic lattice structure, and
nearest-neighbour exchange coupling, in a uniform magnetic field. In this
model, the magnetisation ${\bf m}$ is directed along the field ${\bf h,}$
thus one can set ${\bf h=}h{\bf e}_{z}$ and ${\bf m}_{i}{\bf =}m_{i}{\bf e}%
_{z}.$ In the limit of $D\to \infty $ it becomes immaterial what is the
exact number of ``transverse'' components: $D-2$ for the anisotropic model,
Eq.\ (\ref{Defms}), $D-1$ for the isotropic model in field (the case we are
considering here), or $D$ for the isotropic model in zero field. In general,
there are a few components with nonvanishing averages $\left\langle
s_{\alpha }\right\rangle $, and the remaining components called
``transverse'' components, are booked into the correlation functions.

Solving the system of equations above consists in determining ${\bf m}_{i}$
and $s_{ij}$ as functions of $G_{i}$ from the linear equations (\ref{MagEq})
and (\ref{CFEq}), respectively, and inserting these solutions in the
constraint equation (\ref{Constraint}) in order to obtain $G_{i}$. The two
main types of boundary conditions for our problem are free boundary
conditions (fbc) and periodic boundary conditions (pbc). In the case of fbc,
if the summation index $j$ in Eqs.\ (\ref{MagEq}) and (\ref{CFEq}) runs out
of the lattice, the corresponding terms are omitted. In this case, ${\bf m}%
_{i}$ and $G_{i}$ are inhomogeneous and $s_{ij}$ nontrivially depends on
both indices due to boundary effects. In the pbc case the solution becomes
homogeneous and greatly simplifies. Although the model with pbc is
unphysical, it allows for an analytical treatment at all temperatures and
for the study of finite-size effects separately from boundary effects.

One can also introduce a matrix formalism and rewrite Eqs.\ (\ref{MagEq})
and (\ref{CFEq}) for the isotropic model in field as follows 
\begin{equation}  \label{DefMatr}
\sum_j {\cal D}_{ij} m_j = h, \qquad \sum_j {\cal D}_{ij} s_{jl} = \theta
\delta_{il},
\end{equation}
where we have defined the Dyson matrix ${\cal D}_{ij} \equiv
G_i^{-1}\delta_{ij}-\lambda_{ij}$. The solutions of these linear equations
are 
\begin{equation}  \label{MatrSol}
m_i = h \sum_j {\cal D}_{ij}^{-1} , \qquad s_{ij} = \theta {\cal D}%
_{ij}^{-1} ,
\end{equation}
${\cal D}^{-1}$ being the inverse of the Dyson matrix ${\cal D}$.
Substituting these solutions into the constraint equation (\ref{Constraint})
results in a closed system of nonlinear equations for the gap parameter $G_i$
\begin{equation}  \label{DefG}
\bigg (h \sum_j {\cal D}_{ij}^{-1} \bigg )^2 + \theta {\cal D}_{ii}^{-1} = 1.
\end{equation}

The average magnetisation per site defined by 
\begin{equation}
{\bf m=}\frac{1}{{\cal N}}\sum\limits_{i}{\bf m}_{i}  \label{m}
\end{equation}
vanishes for finite-size systems in the absence of magnetic field due to the
Golstone mode corresponding to the global rotation of the magnetisation. On
the other hand, it is clear that at temperatures $\theta \ll 1$ the spins in
the system are aligned with respect to each other and there should exist an 
{\it intrinsic} {\it magnetisation}. The latter is usually defined for
finite-size systems as 
\begin{equation}
M=\sqrt{\left\langle \left( \frac{1}{{\cal N}}\sum\limits_{i}{\bf s}%
_{i}\right) ^{2}\right\rangle }=\sqrt{{\bf m}^{2}+\frac{1}{{\cal N}^{2}}%
\sum\limits_{i,j=1}^{{\cal N}}s_{ij}},  \label{M}
\end{equation}
where the second expression is valid in the limit $D\rightarrow \infty .$
Here ${\bf m}$ and $s_{ij}$ are defined by Eqs.\ (\ref{Defms}) and (\ref{m}%
). One can see that $M\geq m.$ Note that $M$ remains non zero for $h=0$. In
this case in the limit $\theta \rightarrow 0$ one has $s_{ij}=1$ for all $i$
and $j,$ and $M\rightarrow 1.$ For $\theta \rightarrow \infty $ the spins
becomes uncorrelated, $s_{ij}=\delta _{ij},$ and $M\rightarrow 1/\sqrt{{\cal %
N}}.$ In the limit of ${\cal N}\rightarrow \infty ,$ the intrinsic
magnetisation $M$ approaches that of the bulk system.

The applied field suppresses the global-rotation Goldstone mode and thereby
renders the magnetisation ${\bf m}$ of Eq.(\ref{m}) non zero. Therefore, it
is convenient to call the latter the {\it induced magnetisation,} in
contrast with the intrinsic magnetisation $M$. If the field is strong or the
temperature is low, the spins align along the field, the transverse
correlation functions $s_{ij}$ in Eq.\ (\ref{M}) become small, and the
magnitude of the intrinsic magnetisation approaches the induced
magnetisation.

One can establish an important relation between the intrinsic magnetisation $%
M$ and induced magnetisation $m$. To this end, one first obtains from Eqs.\ (%
\ref{MatrSol}) and (\ref{m}) the useful relation 
\begin{equation}
\frac{1}{{\cal N}^{2}}\sum_{ij}s_{ij}=\frac{\theta }{{\cal N}}\frac{m}{h}.
\label{sijm}
\end{equation}
Then, substituting the latter in Eq.\ (\ref{M}) and solving the equation
thus obtained for $m$, $m^{2}+\theta m/({\cal N}h)-M^{2}=0$, yields 
\begin{equation}
m=M\frac{2{\cal N}Mh/\theta }{1+\sqrt{1+(2{\cal N}Mh/\theta )^{2}}}=MB({\cal %
N}MH/T),  \label{mvsM}
\end{equation}
where $B(\xi )=(2\xi /D)/\left[ 1+\sqrt{1+(2\xi /D)^{2}}\right] $ is the
Langevin function for $D\gg 1$. One can find in the literature formulae of
the type $m=M_{s}B({\cal N}M_{s}H/T)$, where $M_{s}$ is usually associated
with the bulk magnetisation at a given temperature (see, e.g., Refs.\ \cite
{fispri85,fispri86}). In our case, Eq.\ (\ref{mvsM}) is exact and $M=M(T,H)$
is explicitly defined by Eq.\ (\ref{M}). For large sizes ${\cal N}$, Eq.\ (%
\ref{mvsM}) describes two distinct field ranges separated at $h\sim h_{v}$
where 
\begin{equation}
h_{v}\equiv \frac{\theta }{{\cal N}M_{0}(\theta )},\qquad M_{0}(\theta
)\equiv M(\theta ,h=0).  \label{Defhstar}
\end{equation}
In the range $h\,\raisebox{0.35ex}{$<$}\hspace{-1.7ex}\raisebox{-0.65ex}{$%
\sim$}\,h_{v}$, the total magnetic moment of the system is disoriented by
thermal fluctuations, and the induced magnetisation $m$ is lower than the
intrinsic magnetisation $M$. In the range $h\,\raisebox{0.35ex}{$>$}\hspace{%
-1.7ex}\raisebox{-0.65ex}{$\sim$}\,h_{v}$, the total magnetic moment is
oriented by the field, $m$ approaches $M$, and both the latter further
increase with field towards saturation ($m=M=1$) due to the suppression of
spin waves in the system. This scenario is quite general and inherent to all 
$O(D)$ models, as was shown on phenomenological grounds in Ref.\ \cite
{fispri85}. Early Monte Carlo simulations of Ref.\ \cite{wil74} for
isotropic Heisenberg systems of spherical shape confirm Eq.\ (\ref{mvsM})
within statistical errors, although the accuracy is not high enough to
decide whether this relation is exact.

One can also consider local magnetisations. The local induced magnetisation
is simply the vector ${\bf m}_{i},$ while the local intrinsic magnetisation
can be defined as follows 
\begin{equation}
M_{i}=\frac{1}{M}\left\langle {\bf s}_{i}\cdot \frac{1}{{\cal N}}%
\sum\limits_{j}{\bf s}_{j}\right\rangle =\frac{1}{{\cal N}M}%
\sum\limits_{j=1}^{{\cal N}}\left( s_{ij}+{\bf m}_{i}\cdot {\bf m}%
_{j}\right)  \label{Mi}
\end{equation}
One can check the identity $(1/{\cal N})\sum_{i}M_{i}=M$ showing the
self-consistency of the definition given above.

We start by considering the pbc model in the next section.

\section{Systems with periodic boundary conditions}

\label{pbc}

In this case, as was said above, the system becomes homogeneous, that is $%
G_{i}=G$, ${\bf m}_{i}={\bf m}$, and the correlation functions can be found
by performing a Fourier transformation of the type 
\begin{equation}
F_{i}=\frac{1}{{\cal N}}\sum_{{\bf k}}e^{-i{\bf kr}_{i}}F_{{\bf k}},\qquad
F_{{\bf k}}=\sum_{i}e^{i{\bf kr}_{i}}F_{i},  \label{FourierDef}
\end{equation}
where for a cube (${\cal N}=N^{3}$) one has 
\begin{equation}
k_{\alpha }=2\pi n_{\alpha }/N,\qquad n_{\alpha }=0,1,\ldots ,N-1,\qquad
\alpha =x,y,z.  \label{QuantPBC}
\end{equation}
This results in 
\begin{equation}
s_{ii}=\frac{1}{{\cal N}}\sum\limits_{{\bf k}}s_{{\bf k}}=\theta GP_{N}(G),
\label{Constrpbc}
\end{equation}
where 
\begin{equation}
P_{N}(G)=\frac{1}{{\cal N}(1-G)}+\tilde{P}_{N}(G),\qquad \tilde{P}%
_{N}(G)\equiv \frac{1}{{\cal N}}\sum_{{\bf k}}{}^{^{\prime }}\frac{1}{%
1-G\lambda _{{\bf k}}}  \label{PNPNtilde}
\end{equation}
and $\lambda _{{\bf k}}=J_{{\bf k}}/J_{0}$. In the lattice Green function $%
P_{N}(G)$ the contributions of the would-be Goldstone mode with ${\bf k=0}$
and other modes have been separated from each other (hence the prime on the
sum in $\tilde{P}_{N}(G)$). In the bulk limit $N\rightarrow \infty $ for the
sc lattice 
\begin{equation}
P_{\infty }(G)\equiv P(G)\cong \left\{ 
\begin{array}{ll}
1+G^{2}/(2d), & G\ll 1 \\ 
W-c_{0}\sqrt{1-G}, & 1-G\ll 1,
\end{array}
\right.  \label{PLims}
\end{equation}
where $d=3$, the Watson integral $W=1.51639$, and $c_{0}=(2/\pi )(3/2)^{3/2}$%
. The difference between the sum and the integral, 
\begin{equation}
W_{N}^{{\rm (pbc)}}=\frac{1}{{\cal N}}\sum_{{\bf k}}{}^{^{\prime }}\frac{1}{%
1-\lambda _{{\bf k}}}\qquad {\rm and}\qquad W=\int \!\!\!\frac{d^{3}{\bf k}}{%
(2\pi )^{3}}\frac{1}{1-\lambda _{{\bf k}}},  \label{WNpbc}
\end{equation}
describes the finite-size effect and for the sc lattice, $\lambda _{{\bf k}%
}=(\cos k_{x}+\cos k_{y}+\cos k_{z})/d$, $d=3,$ behaves as $1/N$ \cite
{fispri86} 
\begin{equation}
\Delta _{N}^{{\rm (pbc)}}\equiv \frac{W_{N}^{{\rm (pbc)}}-W}{W}\cong -\frac{%
0.90}{N},\qquad N\gg 1.  \label{DeltaPBC}
\end{equation}
The easiest way to obtain the coefficient in this formula is to plot $\Delta
_{N}^{{\rm (pbc)}}$ vs $1/N$ (see Fig.\ \ref{sef_del}). For finite $N$, $%
\tilde{P}_{N}(G)$ becomes regular at $G=1$. In the large-$N$ limit for $%
1-G\ll k_{{\rm min}}^{2}\sim 1/N^{2}$, where $k_{{\rm min}}\sim 1/N$ is the
minimal wave vector in a finite-size system, one has 
\begin{equation}
\tilde{P}_{N}(G)\cong W_{N}^{{\rm (pbc)}}-c_{N}N(1-G),\qquad c_{N}=\frac{%
(2d)^{2}}{(2\pi )^{4}}\sum_{n_{x},n_{y},n_{z}=-\infty }^{\infty
}\!\!\!\!\!\!\!\!\!\!\!\!{}^{^{\prime }}\;\;\;\;\;\;\frac{36}{%
(n_{x}^{2}+n_{y}^{2}+n_{z}^{2})^{2}}\simeq 0.382.  \label{CPBC}
\end{equation}
For larger $1-G$, the square-root singularity in Eq.\ (\ref{PLims}) is
restored. More precisely, one obtains, 
\begin{equation}
\tilde{P}_{N}(G)\cong W_{N}^{{\rm (pbc)}}-\left\{ 
\begin{array}{l}
c_{N}N(1-G),\quad 1-G\ll 1/N^{2}, \\ 
c_{0}\sqrt{1-G},\quad 1-G\gg 1/N^{2}.
\end{array}
\right.  \label{limitsPNtilde}
\end{equation}

Using $\sum_{j}\lambda _{ij}=1$ in the second of Eqs.\ (\ref{MagEq}), one
arrives in the isotropic case $\eta =1$, at the system of equations 
\begin{equation}
m=\frac{hG}{1-G},\quad m^{2}+\theta GP_{N}(G)=1.  \label{pbcEqs}
\end{equation}
The solution $G(\theta )$ of these equations decreases with increasing
temperature, and at high temperatures, the leading asymptote is $G=1/\theta $%
. For the bulk system at low temperatures or high fields, $G$ tends to $%
1/(1+h)$. For finite-size systems, even in zero field, the Goldstone-mode
contribution $1/[{\cal N}(1-G)]$ in Eq. (\ref{PNPNtilde}) makes $G$ smaller
than unity, $1-G\cong \theta /{\cal N}$ at $\theta \ll 1$, which means that
there is a gap in the correlation function $s_{{\bf q}}$. As a consequence,
the magnetisation $m$ in Eq.\ (\ref{pbcEqs}) vanishes for $h\rightarrow 0$
for any finite ${\cal N}$. The same happens for low-dimensional systems $%
d\leq 2$, even in the bulk limit. In contrast, for three-dimensional bulk
systems in zero field, $G$ remains equal to 1 and the gap in the correlation
function closes for $\theta \leq \theta _{c}=1/W$. This is the reason to
call $G$ the gap parameter. Below $\theta _{c}$ the spontaneous bulk
magnetisation is given by 
\begin{equation}
m_{b}=\sqrt{1-\theta /\theta _{c}}  \label{mBulk}
\end{equation}

The intrinsic magnetisation of Eqs.\ (\ref{M}) can with the help of Eqs.\ (%
\ref{sijm}) and the first of Eqs. (\ref{PNPNtilde}) be rewritten as 
\begin{equation}
M=\sqrt{m^{2}+\frac{\theta G}{{\cal N}(1-G)}}=\sqrt{1-\theta G\tilde{P}%
_{N}(G)},  \label{MvsG}
\end{equation}
where the last equality was inferred from the second of Eqs.\ (\ref{pbcEqs}%
). Note that $m$ and $M$ are related by Eq.\ (\ref{mvsM}) which is valid for
all types of boundary conditions. Under a very small magnetic field, we can
write 
\[
\frac{m}{M}=B({\cal N}MH/T)\cong \frac{{\cal N}MH}{T}\equiv \frac{H}{H_{V}}%
\cong 1,
\]
where $H_{V}\equiv \frac{TD}{{\cal N}M}$ is the field at which the global
rotation of the particle's magnetic moment is suppressed. The corresponding
reduced field $h_{v}\equiv \frac{H}{J_{0}}$ was defined in Eq. (\ref
{Defhstar}). Here $1-G$ can be found from the first of Eqs.\ (\ref{pbcEqs})
and Eq.\ (\ref{mvsM}) as 
\begin{equation}
1-G\cong h/m\cong (h_{v}/2)[1+\sqrt{1+(2h/h_{v})^{2}}].  \label{GLowH}
\end{equation}
With the help of these equations, the dependence $M=M(h,\theta )$ can be
established perturbatively in different regions of the parameters. In
particular, below $\theta _{c}$ for ${\cal N}\gg 1$ and $h\ll h_{v},$ one
can use Eq.\ (\ref{limitsPNtilde}) with $1-G\cong h_{v}[1+(h/h_{v})^{2}]$,
which results in 
\begin{equation}
M\cong m_{b}+\frac{\theta }{2m_{b}}\left( -W\Delta _{N}^{{\rm (pbc)}}\right)
+\frac{N^{4}c_{N}}{2}h^{2},\qquad h\ll h_{v}.  \label{MLowLowH}
\end{equation}
The last term in this expression cannot be obtained from the naive spin-wave
theory.

On the other hand, in high fields, there is a crossover to the bulk
spin-wave singularity at the field $H_{S}\equiv \frac{\pi ^{2}}{2d}\frac{%
J_{0}}{{\cal N}^{2/3}}$ which is much larger than $H_{V}.$ Indeed, for $h\gg
h_{s}$ with 
\begin{equation}
h_{s}\equiv \frac{H_{S}}{J_{0}}=\frac{\pi ^{2}}{2d}\frac{1}{N^{2}},
\label{hs}
\end{equation}
Eqs. (\ref{MvsG}), together with the second line of Eq. (\ref{limitsPNtilde}%
), leads to the following high-field behaviour of the magnetisation 
\begin{equation}
M\cong m_{b}+\frac{\theta }{2m_{b}}\left( -W\Delta _{N}^{{\rm (pbc)}}+c_{0}%
\sqrt{h}\right) ,\quad \quad h\gg h_{s}  \label{h>>hs}
\end{equation}
which shows the well known singular $\sqrt{h}$ spin-wave correction to the
magnetisation in three dimensions.

For $h_{v}\ll h\ll h_{s},$ Eq. (\ref{MvsG}) and first line of Eq. (\ref
{limitsPNtilde}) yield the linear field behaviour of the magnetisation as a
function 
\begin{equation}
M\cong m_{b}+\frac{\theta }{2m_{b}}\left( -W\Delta _{N}^{{\rm (pbc)}%
}+Nc_{N}h\right) ,\quad \quad h_{v}\ll h\ll h_{s}.  \label{hv<h<hs}
\end{equation}

Therefore, we obtain two crossovers in field, one at $h_{v}$ given by Eq.\ (%
\ref{Defhstar}) between the quadratic and linear behaviour, and another one
at $h_{s}$ defined in Eq. (\ref{hs}) between the linear and square-root
behaviour for the magnetisation.

Note that both of these crossovers occur for the induced magnetisation $m$,
Eq.\ (\ref{mvsM}), as well as the intrinsic magnetisation $M$, Eq.\ (\ref
{MvsG}). At low temperatures in zero field $M$ deviates from 1 according to
the law 
\begin{equation}
M\cong 1-\theta W_{N}^{{\rm (pbc)}}/2,  \label{MLowTpbc}
\end{equation}
where the coefficient in the linear-$\theta $ term is smaller than in the
bulk [see Eq.\ (\ref{DeltaPBC})].

It should be noted that in his early work on thin magnetic films D\"{o}ring 
\cite{doe61} used expressions of the type of Eq.\ (\ref{MLowTpbc}), where
the dangerous ${\bf k}=0$ mode was excluded on physical grounds. This
intuitive approach was applied to clusters of quantum spins in Ref.\ \cite
{henlinlin93} using the numerical solution for the energy levels in the
linear spin-wave approximation. Whereas the results of these works for the
temperature dependence of the magnetisation are reasonable if one specifies
the meaning of the ``magnetisation'' as the {\em intrinsic} magnetisation in
zero field, Eq.\ (\ref{mvsM}) relating the intrinsic and induced
magnetisations with each other and describing global fluctuations cannot be
obtained if one just neglects the ${\bf k}=0$ mode.

In spatial dimensions $d\leq 2$, $P(G)$ diverges for $G\rightarrow 1$, which
rules out the long-range order in bulk systems. For any finite-size system,
however, $M\to 1$ at low temperatures. In particular, in two dimensions one
has $W_N^{{\rm (pbc)}} \cong 8\pi\ln N + {\rm const}$, which results in 
\begin{equation}  \label{MSW2d}
M \cong 1 - \theta(4\pi\ln N + {\rm const}), \qquad N \gg 1, \qquad \theta
\ll 1.
\end{equation}

For the fields $h\gg 1,\theta $ one has, again, $m\cong 1$ , $G\cong
1/(1+h)\ll 1$, $P(G)\cong 1,$ and thus using the first of Eqs.\ (\ref{pbcEqs}%
) and the second equality in Eq.\ (\ref{MvsG}) one obtains 
\begin{equation}
M\cong 1-\frac{\theta }{2(1+h)}.  \label{MHT}
\end{equation}
Here keeping 1 in the denominator improves the asymptotic behavior for
moderately high values of $h.$

For the model with free boundary conditions, one cannot obtain a general
analytical solution, and in the main range of temperatures and fields the
problem has to be solved numerically. Analitycal solutions exist, however,
for high and low temperatures and for high fields. Before proceeding with
the numerical solution of our problem, we will consider in the next section
the most interesting analytical solution of the fbc model for $\theta \ll 1$.

\section{Boundary effects at low temperature}

\label{LowT}

At zero field the induced magnetisation ${\bf m}$ is zero, but in the limit
of zero temperature all spins in the system become strongly correlated with
each other: $s_{ij}\rightarrow 1$. For $\theta \ll 1$ one can search for the
solution of Eqs.\ (\ref{CFEq}) and (\ref{Constraint}) in the form 
\begin{equation}
s_{ij}\cong 1-\delta s_{ij},\qquad G_{i}\cong G_{i}^{(0)}-\delta G_{i},
\label{LTAnsats}
\end{equation}
where $G_{i}^{(0)}$ is the zero-temperature value of $G_{i},$ $\delta s_{ij}$
and $\delta G_{i}$ are small corrections. The zeroth order of Eq.\ (\ref
{CFEq}) becomes 
\begin{equation}
1=G_{i}^{(0)}\sum_{j}\lambda _{ij}  \label{Gi0Eq}
\end{equation}
which determines $G_{i}^{(0)}$. If the site $i$ is not near the boundary so
that all sites $j$ are within the system, then the sum above and thus $%
G_{i}^{(0)}$ are equal to one. If $i$ is near the boundary, some of the
values of $j$ run out of the system and are not counted, thus $G_{i}^{(0)}$
increases. For the sc lattice one has 
\begin{equation}
\frac{1}{G_{i}^{(0)}}=1-\frac{1}{6}(\delta _{i_{x},1}+\delta
_{i_{x},N}+\delta _{i_{y},1}+\delta _{i_{y},N}+\delta _{i_{z},1}+\delta
_{i_{z},N}),  \label{Gi0}
\end{equation}
where $i\equiv \{i_{x},i_{y},i_{z}\}$. That is, $G_{i}^{(0)}=6/5$ on the
faces, $G_{i}^{(0)}=3/2$ on the edges, and $G_{i}^{(0)}=2$ at the corners.

To first order in $\theta ,$ Eqs.\ (\ref{Constraint}) and (\ref{DefMatr})
become 
\begin{eqnarray}
&&\delta s_{ii}=0,  \nonumber  \label{LinearizedEqs} \\
&&\sum_{j}{\cal D}_{ij}^{(0)}\delta s_{jl}=-\theta \delta _{il}+\delta G_{i}/%
\big(G_{i}^{(0)}\big)^{2},
\end{eqnarray}
where ${\cal D}_{ij}^{(0)}$ is the Dyson matrix defined in Eq.\ (\ref
{DefMatr}) taken at zero temperature. The solution of the linear problem
above has the form 
\begin{eqnarray}
&&\delta G_{i}=\theta \big(G_{i}^{(0)}\big)^{2}\sum_{j}{\cal D}_{ij}^{(0)}%
{\cal D}_{jj}^{-1(0)}  \nonumber  \label{LinSol} \\
&&\delta s_{ij}=\theta [{\cal D}_{ii}^{-1(0)}-{\cal D}_{ij}^{-1(0)}],
\end{eqnarray}
where ${\cal D}_{ij}^{-1(0)}$ are the matrix elements of the inverse Dyson
matrix ${\cal D}^{(0)}$.

The latter equations can be used to calculate different observables. In
particular, the intrinsic magnetisation $M$ of Eq.\ (\ref{M}) becomes 
\begin{equation}
M\cong 1-\frac{\theta }{2{\cal N}}\sum_{ij}{\cal D}_{ij}^{-1(0)}\left(
\delta _{ij}-\frac{1}{{\cal N}}\right) .  \label{MLT}
\end{equation}
To evaluate the coefficient in the linear-$\theta $ term, it is convenient
to expand ${\cal D}_{ij}^{-1(0)}$ over the set of eigenfunctions $F_{{\bf k}%
i}$ of the Dyson matrix ${\cal D}_{ij}^{(0)}$: 
\begin{equation}
{\cal D}_{ij}^{-1(0)}=\sum_{{\bf k}}\frac{F_{{\bf k}i}F_{{\bf k}j}}{\mu _{%
{\bf k}}},  \label{DinvExp}
\end{equation}
where the eigenfunctions satisfy 
\begin{equation}
\sum_{i}F_{{\bf k}i}{\cal D}_{ij}^{(0)}=\mu _{{\bf k}}F_{{\bf k}j},\qquad
\sum_{i}F_{{\bf k}i}F_{{\bf k}^{\prime }i}=\delta _{{\bf kk^{\prime }}}.
\label{FEq}
\end{equation}
The latter can be found exactly since the matrix ${\cal D}_{ij}^{(0)}$ can
be represented as a sum of three matrices: 
\begin{eqnarray}
&&{\cal D}_{ij}^{(0)}={\cal D}_{ij}^{x(0)}+{\cal D}_{ij}^{y(0)}+{\cal D}%
_{ij}^{z(0)},  \nonumber  \label{DDecomp} \\
&&\!\!\!\!\!\!\!\!\!\!{\cal D}_{ij}^{x(0)}=-(1/6)[\delta
_{i_{x},j_{x}-1}+\delta _{i_{x},j_{x}+1}-\delta _{i_{x},j_{x}}(2-\delta
_{j_{x},1}-\delta _{j_{x},N})]\delta _{i_{y},j_{y}}\delta _{i_{z},j_{z}},
\end{eqnarray}
etc., each of them acting as a discrete Laplace operator along the
respective coordinate. Thus the variables separate 
\begin{equation}
F_{{\bf k}i}=f_{i_{x},k_{x}}\times f_{i_{y},k_{y}}\times
f_{i_{z},k_{z}},\qquad \mu _{{\bf k}}=\tilde{\mu}_{k_{x}}+\tilde{\mu}%
_{k_{y}}+\tilde{\mu}_{k_{z}},  \label{SepVars}
\end{equation}
and one obtains $\mu _{{\bf k}}=1-\lambda _{{\bf k}}$ and 
\begin{equation}
f_{i_{\alpha },k_{\alpha }}\propto \cos [(i_{\alpha }-1/2)k_{\alpha
}],\qquad k_{\alpha }=\pi n_{\alpha }/N,\qquad n_{\alpha }=0,1,\ldots ,N-1
\label{SepSol}
\end{equation}
with $\alpha =z,y,z$.

\begin{figure}[t]
\unitlength1cm 
\begin{picture}(15,7.5)
\centerline{\psfig{file=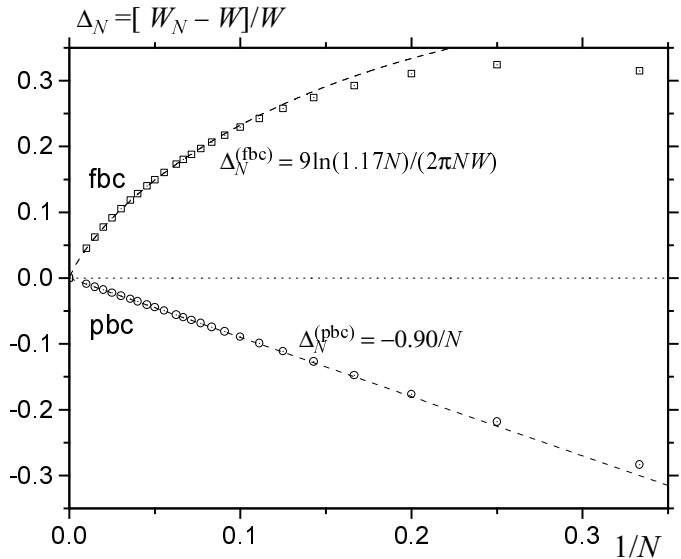,angle=-90,width=10cm}}
\end{picture}
\caption{Lattice sums $W_N$ for the systems of the cubic form with free and
periodic boundary conditions. $W=1.51639$ is the bulk value for the sc
lattice. }
\label{sef_del}
\end{figure}

Now, after substituting Eq.\ (\ref{DinvExp}) in Eq.\ (\ref{MLT}), one
obtains for $\theta \ll 1$ 
\begin{equation}
M\cong 1-\frac{\theta }{2}W_{N}^{{\rm (fbc)}},\qquad W_{N}^{{\rm (fbc)}}=%
\frac{1}{{\cal N}}\sum_{{\bf k}}{}^{^{\prime }}\frac{1}{1-\lambda _{{\bf k}}}
\label{WNfbc}
\end{equation}
[cf. Eq.\ (\ref{MLowTpbc})]. It is interesting to note that $W_{N}^{{\rm %
(fbc)}}$ differs from $W_{N}^{{\rm (pbc)}}$ of Eq.\ (\ref{WNpbc}) only by
the definition of the discrete wave vectors, Eqs.\ (\ref{QuantPBC}) and (\ref
{SepSol}), which results from different boundary conditions at the surfaces.
In contrast to $W_{N}^{{\rm (pbc)}}$, the value of $W_{N}^{{\rm (fbc)}}$
exceeds the bulk value $W$, because the spins near the surface fluctuate
more strongly at finite temperatures. Moreover, it can be shown that in the
isotropic semi-infinite system the magnetisation profile is long ranged and
decays as $1/l$, where $l$ is the distance from the surface. Integrating
this profile over the particle's volume gives $\Delta _{N}^{{\rm (fbc)}%
}=(W_{N}^{{\rm (fbc)}}-W)/W\propto \ln (N)/N$, i.e., the difference from the
bulk value for the fbc system decreases more slowly than $\Delta _{N}^{{\rm %
(pbc)}}$ at large $N$. To obtain this result explicitly, one should bring $%
W_{N}^{{\rm (fbc)}}$ and $W_{N}^{{\rm (pbc)}}$ to the forms that are
comparable with each other. To this end, we write $W_{2N}^{{\rm (pbc)}}$ in
the form 
\begin{equation}
W_{2N}^{{\rm (pbc)}}=\frac{1}{8{\cal N}}\sum_{{\bf k}}{}^{^{\prime }}\frac{1%
}{1-\lambda _{{\bf k}}},\qquad k_{\alpha }=\frac{\pi n_{\alpha }}{N},\qquad
n_{\alpha }=-N,\ldots ,N-1  \label{W2Npbc}
\end{equation}
and complete the range of ${\bf k}$ in Eq.\ (\ref{WNfbc}) to that of the
formula above. This results in the exact relation 
\begin{eqnarray}
&&W_{N}^{{\rm (fbc)}}=W_{2N}^{{\rm (pbc)}}+\frac{9}{16{\cal N}}%
\sum_{k_{x},k_{y}}{}^{^{\prime }}\frac{1}{(1-\lambda _{{\bf k}%
}^{(2)})(2-\lambda _{{\bf k}}^{(2)})}  \nonumber  \label{pbcfbc} \\
&&\qquad {}+\frac{9}{{\cal N}}\sum_{k}{}^{^{\prime }}\frac{1}{(1-\lambda
_{k}^{(1)})(3-\lambda _{k}^{(1)})(5-\lambda _{k}^{(1)})}-\frac{7}{4{\cal N}},
\end{eqnarray}
where $\lambda _{{\bf k}}^{(2)}\equiv (\cos k_{x}+\cos k_{y})/2$, $\lambda
_{k}^{(1)}\equiv \cos k$, and the additional terms arise because of the
double counting on different faces, edges, and corners of the Brillouin zone
as the completion is done. These terms can be loosely interpreted as
contributions from faces, edges, and corners of the cube. One can see that
in the large-$N$ limit the face contributions are the most important ones
and they behave as $\ln (N)/N$, whereas the edge contributions $\sim 1/N$
and the corner contributions $\sim 1/N^{3}$. The result for $\Delta _{N}^{%
{\rm (fbc)}}$ has the form 
\begin{equation}
\Delta _{N}^{{\rm (fbc)}}\equiv \frac{W_{N}^{{\rm (fbc)}}-W}{W}\cong \frac{%
9\ln (1.17N)}{2\pi NW},\qquad N\gg 1,  \label{DeltaFBC}
\end{equation}
cf. Eq.\ (\ref{DeltaPBC}). The next-to-log term in Eq.\ (\ref{DeltaFBC}) is
more difficult to obtain analytically; the best way is to fit the result of
the direct numerical calculation of Eq.\ (\ref{WNfbc}) (see Fig.\ \ref
{sef_del}).

\section{Numerical results}

\label{numres}

Here we present our numerical results for the temperature and field
dependence of the intrinsic magnetisation $M$ and induced magnetisation $m$
for small magnetic systems with simple cubic lattice and cubic shape,
subject to free and periodic boundary conditions (fbc and pbc). As was
mentioned in Sec.\ \ref{basic}, the numerical method for solving the $D\to
\infty $ model consists in obtaining the correlation function $s_{ij}$ and
magnetisation $m_{i}$ from the linear equations (\ref{DefMatr}),
substituting them into the constraint equation (\ref{Constraint}), and
solving the resulting nonlinear equation for the gap parameter $G_{i}$. On
the first step we use a linear-equation solver based on the sparce-matrix
storage appropriate for the Dyson matrix ${\cal D}$, which is defined by
Eq.\ (\ref{DefMatr}), in our case of nearest-neighbour interactions. The
nonlinear problem is tackled using a nonlinear-equation solver of the
Newton-Raphson kind. For the present geometry, symmetry arguments have been
used to reduce the number of unknowns and thereby to increase the computing
performance. The relevant subset of sites corresponding to different values
of $G_{i}$ is obtained by taking 1/8 of the cube with a subsequent reduction
by a factor of about 3! using permutations of the coordinates $x$, $y$, and $%
z$. This allows us to reduce the number of unknowns by a factor of about $%
8\times 3!=48$, this estimation becomes asymptotically exact for ${\cal N}%
\gg 1$. For the square lattice the reduction factor is about $4\times 2!=8$.
All computations have been performed on a Pentium III/450 MHz of 128 MB RAM.

\begin{figure}[t]
\unitlength1cm 
\begin{picture}(15,7.5)
\centerline{\psfig{file=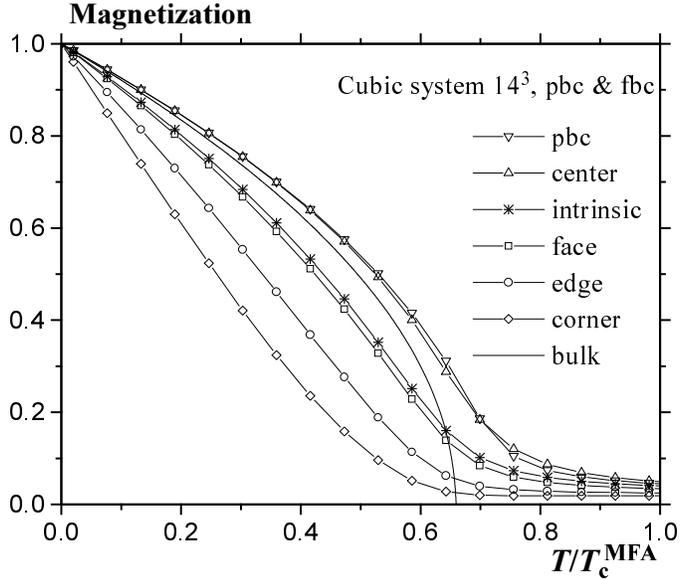,angle=-90,width=10cm}}
\end{picture}\label{sef_t3d}
\caption{Temperature dependence of the intrinsic magnetization $M$, 
Eq.\ (\protect\ref{M}), and local magnetizations $M_i$, Eq.\ (\protect\ref{Mi}), of the $14^3$
cubic system with free and periodic boundary conditions in zero field.}
\end{figure}

\begin{figure}[t]
\unitlength1cm 
\begin{picture}(15,6.5)
\centerline{\psfig{file=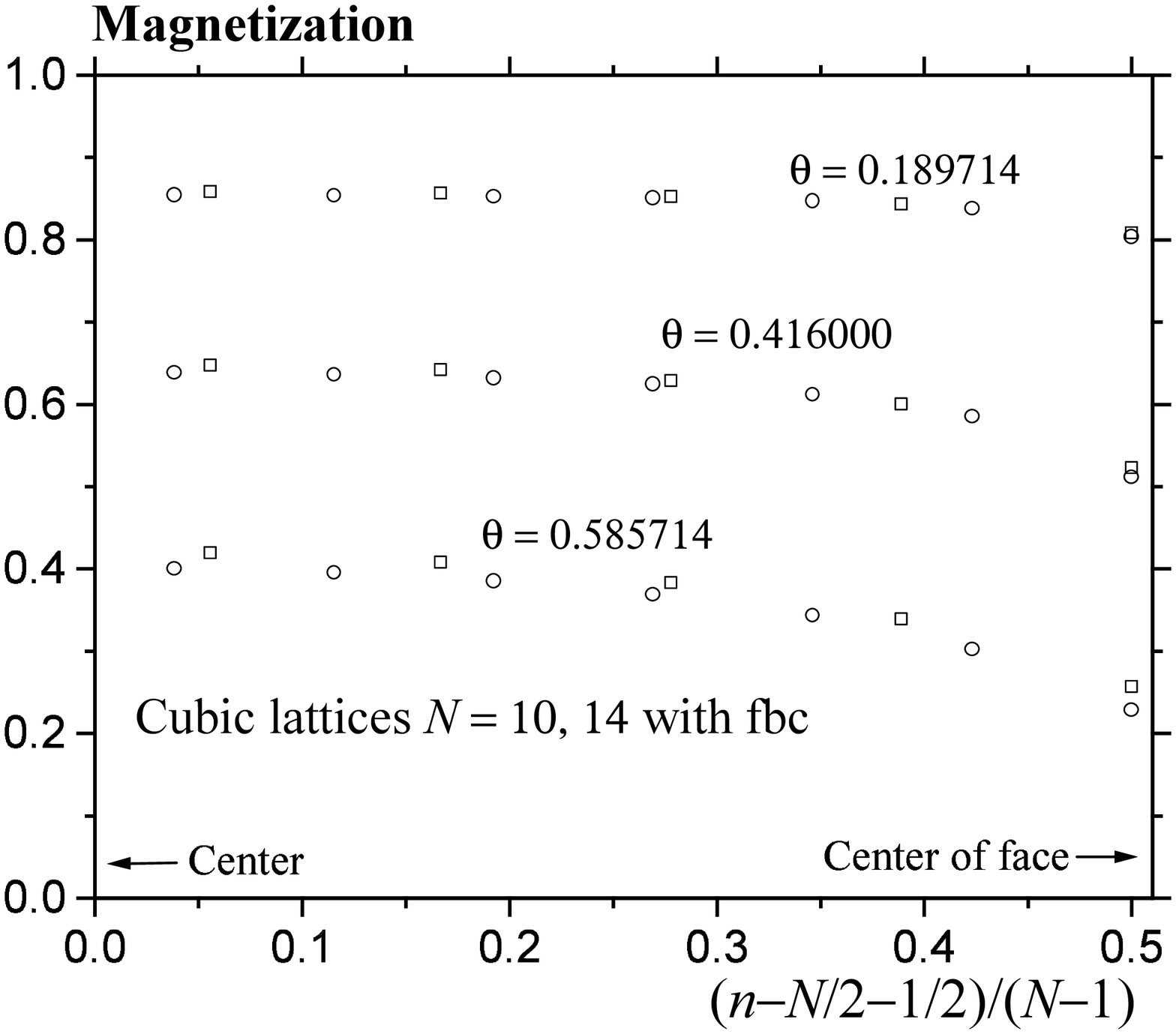,angle=-90,width=10cm}}
\end{picture}
\par
\vspace{0.5cm}
\par
\begin{picture}(15,7)
\centerline{\psfig{file=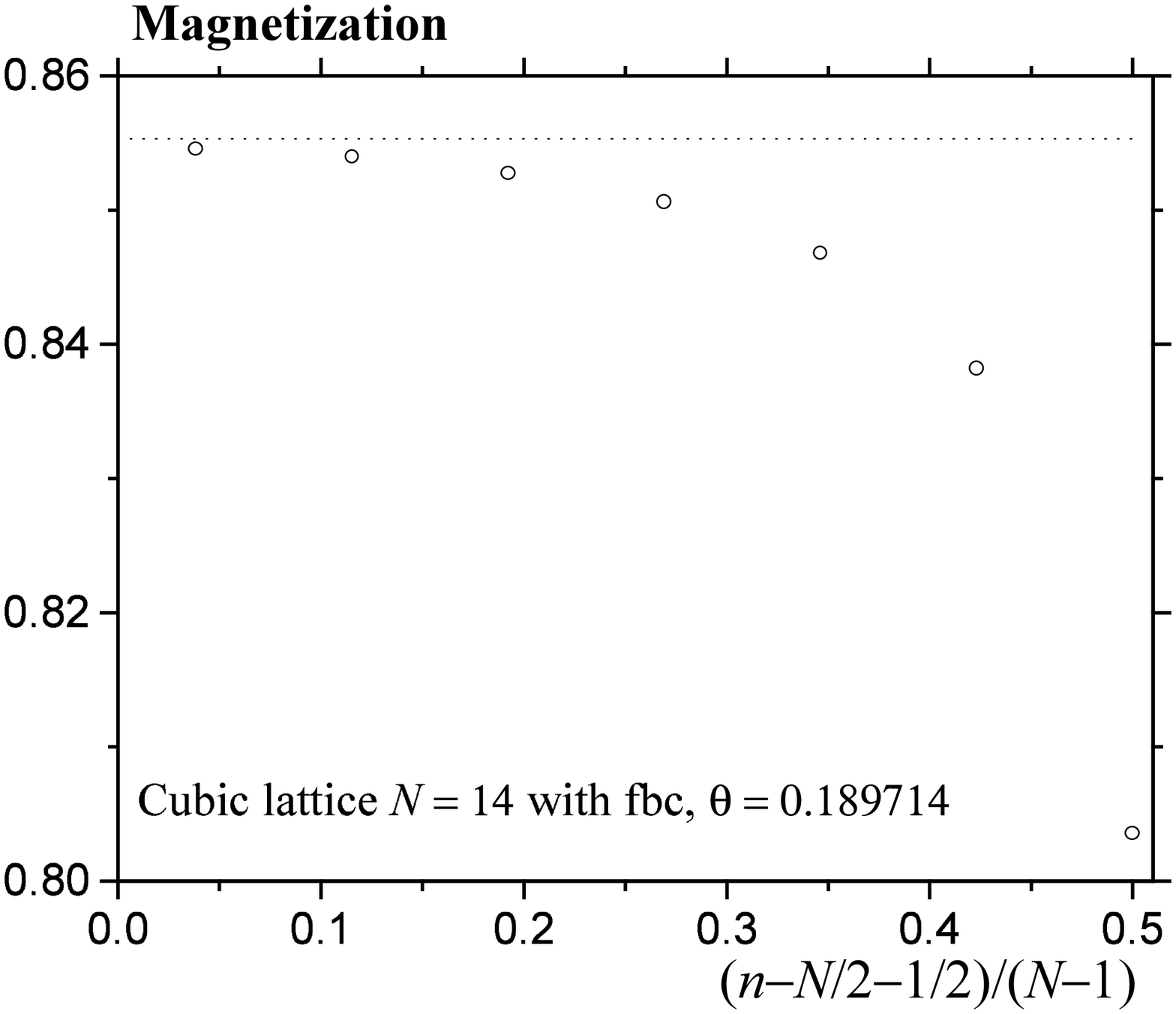,angle=-90,width=10cm}}
\end{picture}
\caption{Magnetization profiles in the direction from the center of the cube
to the center of a face at different temperatures $\theta\equiv T/T_c^{{\rm %
MFA}}$. The magnified lower plot shows the long-range character of the
profile at low temperatures. }
\label{sef_pr}
\end{figure}

\begin{figure}[t]
\unitlength1cm 
\begin{picture}(15,7.5)
\centerline{\psfig{file=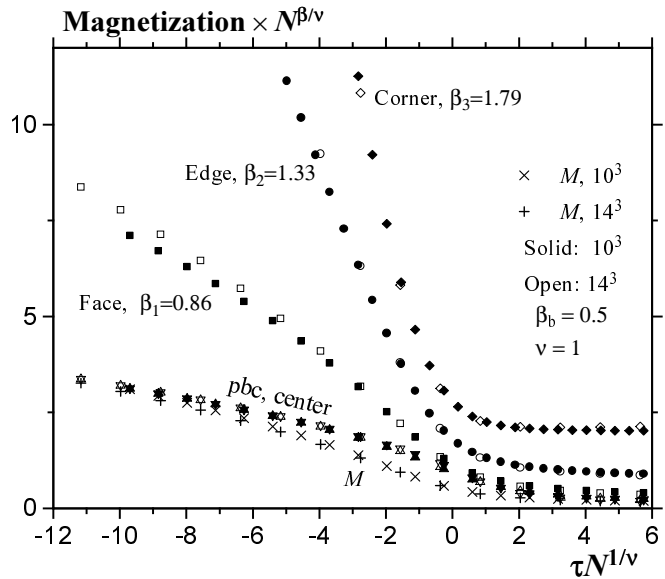,angle=-90,width=10cm}}
\end{picture}
\caption{Finite-size-scaling plot for the magnetization of cubic systems
with linear size $N=10$ and 14.}
\label{sef_s}
\end{figure}

\begin{figure}[t]
\unitlength1cm 
\begin{picture}(15,7.5)
\centerline{\psfig{file=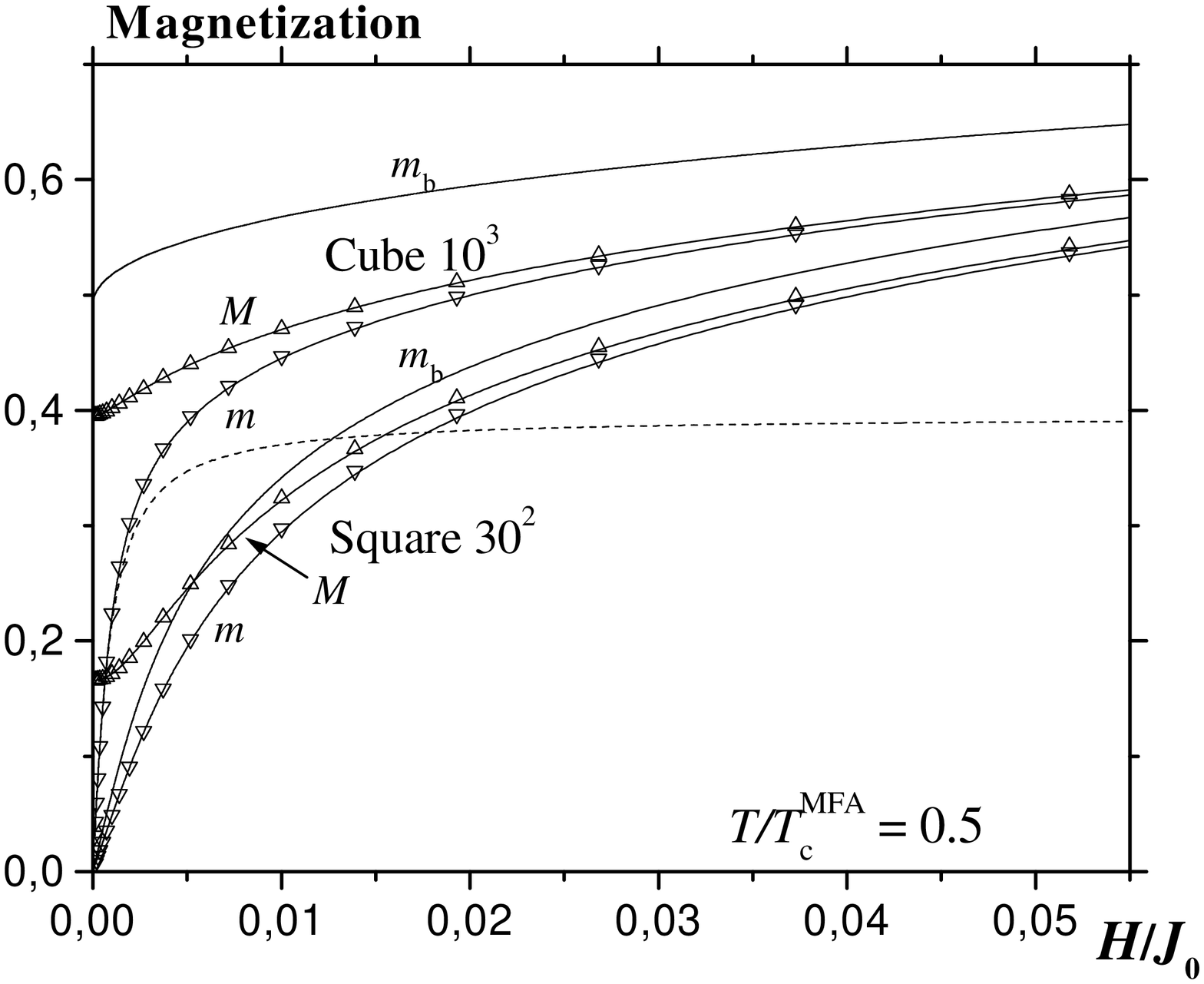,angle=-90,width=10cm}}
\end{picture}
\caption{Field dependence of the intrinsic magnetization $M$ and intrinsic
magnetization $m$ for hypercubic lattices with fbc in three and two
dimensions. Dashed line is a plot of Eq.\ (\protect\ref{mvsM}) in which $M(H,T)$ is
replaced by its zero-field value. Bulk magnetization $m_{{\rm b}}$ in two
and three dimensions is shown by solid lines. }
\label{sef_h}
\end{figure}

Fig.\ \ref{sef_t3d} shows the temperature dependence of the intrinsic
magnetisation $M$, Eq.\ (\ref{M}), and local magnetisations $M_{i}$, Eq.\ (%
\ref{Mi}), of the $14^{3}$ cubic system with free and periodic boundary
conditions in zero field. For periodic boundary conditions, $M$ exceeds the
bulk magnetisation at all temperatures. In particular, at low temperatures
this is in accord with the positive sign of the finite-size correction to
the magnetisation, Eqs.\ (\ref{h>>hs}) and (\ref{DeltaPBC}). The
magnetisation at the center of the cube with free boundary conditions is
rather close to that for the model with pbc in the whole temperature range
and converges with the latter at low temperatures. Local magnetisations at
the center of the faces and edges and those at the corners decrease with
temperature much faster than the magnetisation at the center. This is also
true for the intrinsic magnetisation $M$ which is the average of the local
magnetisation $M_{i}$ over the volume of the system: For the relatively
small size $14^{3}$ the contribution from the boundaries to the average
properties are still substantial. One can see that, in the temperature range
below the bulk critical temperature, $M$ is smaller than the bulk
magnetisation. This means that the boundary effects suppressing $M$ are
stronger than the finite-size effects which lead to the increase of the
latter. This is also seen from the low-temperature expresion for $M$ given
in Eq.\ (\ref{WNfbc}).

Magnetization profiles in the direction from the center of the cube to the
center of a face are shown in Fig.\ \ref{sef_pr} at different temperatures.
One can see that perturbations due to the free boundaries extend deep into
the system. This is a consequence of the Goldstone mode which renders the
correlation length of an isotropic bulk magnet infinite below $T_{c}$. The
latter effect is better seen from the magnified data at the low temperature
(see the lower plot), where the MFA predicts, on the contrary, a fast
approach to a constant magnetisation when moving away from the surfaces \cite
{wil74}. MC simulations of the classical Heisenberg model \cite
{wil74,kacetal99} also show long-range magnetisation profiles. The
difference of the local magnetisations in the center and at a surface point
reaches a maximun in the vicinity of the bulk $T_{c}$ and goes to zero at
low and high temperatures.

Fig.\ \ref{sef_s} indicates that the critical indices for the magnetisation
at the faces, edges, and corners are higher than the bulk critical index $%
\beta =1/2$ for the present $D=\infty $ model. The critical index at the
face $\beta _{1}$ is the mostly studied surface critical index (see, for a
review, Refs.\ \cite{bin83ptcp,die86ptcp}). The exact solution of Bray and
Moore \cite{bramoo77prljpa} for the correlation functions at criticality in
the $D=\infty $ model and application of the scaling arguments yield the
value $\beta _{1}=1$ (see Table II in Ref.\ \cite{bin83ptcp}). Exact values
of the edge and corner magnetisation indices, $\beta _{2}$ and $\beta _{3}$,
seem to be unknown for $D=\infty $. Cardy \cite{car83} used the first-order $%
\varepsilon $-expansion to obtain $\beta _{2}(\alpha )$ for the edge with an
arbitrary angle $\alpha $. For $\alpha =\pi /2$ and $D=\infty $ in three
dimensions the result for the edge critical exponent reads $\beta
_{2}=13/8+O(\varepsilon ^{2})=1.625+O(\varepsilon ^{2})$.

To estimate the magnetisation critical indices in our model we have
performed a finite-size-scaling analysis (see, for a review, Ref.\ \cite
{bin92fss}) assuming the scaling form $M=N^{-\beta /\nu }F_{M}(\tau N^{1/\nu
})$ and plotting in Fig.\ \ref{sef_s} the magnetisation times $N^{\beta /\nu
}$ vs $\tau N^{1/\nu }$. Here $\nu =1$ is the critical index for the
correlation length in the bulk and $\tau \equiv T/T_{c}-1$, where $%
T_{c}=T_{c}^{{\rm MFA}}/W$ is the bulk Curie temperature. Our results for
the systems with $N=10$ and $N=14$ merge into single ``master curves'' for $%
\beta _{1}=0.86$, $\beta _{2}=1.33$, and $\beta _{3}=1.79$, which have been
obtained by fitting $M\propto N^{-\beta /\nu }$ at $T=T_{c}$, i.e., $\theta
=\theta _{c}=1/W$. Note that our value 0.86 for the surface magnetisation
critical index $\beta _{1}$ is substantially lower than the value $\beta
_{1}=1$ following from scaling arguments. This disagreement is probably due
to corrections to scaling which could be pronounced for our insufficiently
large linear sizes $N=10$ and 14. A more efficient way for obtaining an
accurate value of $\beta _{1}$ is to perform a similar analysis for the
semi-infinite model. The latter was considered analytically and numerically
for $T\geq T_{c}$ and $H=0$ in Ref.\ \cite{gar98pre}. We also mention the
Monte Carlo simulations of the Ising model \cite{plesel98} which yield $%
\beta _{1}=0.80$, $\beta _{2}=1.28$, and $\beta _{3}=1.77$.

The field dependence of $M$ and $m$ at fixed temperature, as obtained from
the numerical solution of Eqs.\ (\ref{MagEq}) -- (\ref{Constraint}), for
cubic and square systems is shown in Fig.\ \ref{sef_h}. Naturally the
numerical results for $m$ confirm Eq.\ (\ref{mvsM}) which describes both the
effect of orientation of the system's magnetisation by the field and the
increase of $M$ in field. On the contrary, using the zero-field value of $M$
in Eq.\ (\ref{mvsM}) leads to a poor result for $m$ shown by the dashed
curve for the 10$^{3}$ system in Fig.\ \ref{sef_h}. Here we do see that
there are two distinct field ranges separated by the characteristic field $%
h^{*}$, which were discussed at the end of Sec.\ \ref{basic}. The results of
Fig.\ \ref{sef_h} are in qualitative agreement with those of Fig.\ 1 in
Ref.\ \cite{fispri85} showing the field dependence of $\chi =\partial
m/\partial h$ obtained from phenomenological arguments. We would like to
stress, however, that Eq.\ (\ref{mvsM}) which is {\em exact} in the $%
D=\infty $ model, has not yet been checked for systems with a finite number
of spin components $D$ and arbitrary size ${\cal N}$. The field dependence
of the particle's magnetisation similar to that shown in Fig.\ \ref{sef_h}
for the cubic system was experimentally obtained for ultrafine cobalt
particles in recent Ref.\ \cite{resetal98}, as well as in a number of
previous experiments.

The curves for the square system in Fig.\ \ref{sef_h} illustrate the fact
that in two dimensions thermal fluctuations are much stronger than in 3$d$,
which leads to lower values of both $M$ and $m$ at the same temperature. The
bulk magnetisation $m_{{\rm b}}$ in two dimensions vanishes at zero field
and it thus goes below the intrinsic magnetisation $M$ in the low-field
region.

\section{Conclusion}

\label{conclusion}

In this paper, we have studied finite-size and boundary effects in small
ferromagnetic systems with free boundaries within the $D=\infty $ classical
vector model. This model, while sacrificing about 5\% in the overall
accuracy when used as a substitute for the Heisenberg model in three
dimensions, is exactly solvable and describes relevant physical features
related to the Goldstone modes, such as the absence of ordering in two
dimensions for isotropic models, long-range magnetisation profiles, etc. The 
$D=\infty $ model is clearly superior to the mean field approximation which
ignores these important effects.

An important result of our work is the exact relation in Eq.\ (\ref{mvsM})
beetween the intrinsic magnetisation $M(H,T)$ and induced magnetisation $%
m(H,T)$. It would be interesting to check whether this plausible relation
holds for a more realistic Heisenberg model. In the latter case, it should
be at least a good approximation.

Whereas the finite-size effects, which occur in a pure form in systems with
periodic boundary conditions, lead to the insrease of the magnetisation with
respect to the bulk value, boundary effects in systems with free boundaries
work in the opposite direction. Since the magnetisation reduction caused by
the boundaries extends over the whole system (i.e., the magnetisation
profiles are long ranged), the boundary effect overweighs the finite-size
effect by a large logarithmic contribution (see Fig.\ \ref{sef_del}), and
the resulting intrinsic magnetisation $M$ is lower than the bulk
magnetisation below the bulk critical temperature. The latter effect and the
long-range magnetisation profiles are illustrated in Figs.\ \ref{sef_t3d}
and \ref{sef_pr}.

We have also attempted to estimate the surface magnetisation critical
indices for the faces, edges, and corners of the cube from the scaling
analysis of the data for cubic systems with linear sizes $N=10$ and $N=14$.
It seems that our values for all these indices are too low and the data for
larger sizes are needed. The required computer resources, however, rapidly
increase with the system's linear size $N$, since the theory operates with
the correlation matrices of the size $N^{3}\times N^{3}$.

Field dependences of the particle's induced magnetisation $m$ in Fig. \ref
{sef_h} reflect both the {\em alignment} of the intrinsic magnetic moment $M$
by the field and the {\em increase} of $M$ in field. A combination of these
two effects was observed in many experiments including the recent
publication \cite{resetal98}.

The present approach can be extended to magnetic systems of other shapes
(spherical, cylindrical, etc.), more complicated lattice structures, and
systems with a bulk and surface anisotropy. This will also allow us to
compare with the results of the Monte Carlo simulations of a nanoparticle
(of the maghemite type) studied in \cite{kacetal99}. We are planning to
study these issues in our future work.

\section*{Acknowledgements}

D. A. Garanin is endebted to the Universit\'e de Versailles Saint Quentin
and Laboratoire de Magn\'etisme et d'Optique for the warm hospitability
extended to him during his stay in Versailles in July 1999 and January 2000.
H. Kachkachi thanks Max-Planck-Institut f\"ur Physik komplexer Systeme
Dresden for inviting him for a short stay in September 1999.


\end{document}